\documentclass{PoS}

\usepackage{graphicx}
\usepackage{dcolumn}
\usepackage{multirow}
\usepackage{subfigure}
\usepackage{times,mathptm}
\usepackage{float}
\usepackage{color}
\usepackage{cite}
\usepackage{amsmath,amsfonts}
\usepackage{mathptmx}
\usepackage{mathrsfs}
\usepackage{bbm}
\usepackage{bm}
\usepackage{xfrac}

\newcommand{\bea}{\begin{eqnarray}}
\newcommand{\eea}{\end{eqnarray}} 
\definecolor{orange}{rgb}{1,0.5,0}
\newcommand{\beq}{\begin{equation}}
\newcommand{\eeq}{\end{equation}}

\renewcommand{\a}{\alpha}
\renewcommand{\b}{\beta}

\newcommand{\tr}{\text{Tr}}

\newcommand{\bx}{\mathbf{x}}
\newcommand{\by}{\mathbf{y}}

\newcommand{\vx}{\bx}
\newcommand{\vy}{\by}
\newcommand{\vz}{\mathbf{z}}

\newcommand{\m}{\mu}
\newcommand{\g}{\gamma}

\newcommand{\s}{\sigma}

\newcommand{\N}{{\cal N}}

\newcommand{\vph}{\varphi}
\newcommand{\oh}{\frac{1}{2}}

\newcommand{\dg}{\dagger}
\newcommand{\non}{\nonumber}

\newcommand{\rf}[1]{(\ref{#1})}
\newcommand{\ra}{\rightarrow}
\newcommand{\pa}{\partial}

\title{On the distinction between color confinement, and confinement}

\ShortTitle{S$_\text{c}$ confinement}

\author{\speaker{Jeff Greensite}\thanks{Work supported by the US Department of Energy under Grant No.\ DE-SC0013682.}\\       
        San Francisco State University\\
        E-mail: \email{greensit@sfsu.edu}}

\author{Kazue Matsuyama \\
        San Francisco State University \\
        E-mail: \email{kazuem@sfsu.edu}}

\abstract{The property of color confinement (``C confinement''), meaning that all asymptotic particle states are color neutral, holds not only in QCD, but also in gauge-Higgs theories deep in the Higgs regime. In this talk we describe a new and stronger confinement criterion, separation-of-charge confinement or ``S$_\text{c}$ confinement,'' which is an extension of the Wilson area-law criterion to gauge + matter theories. We will show that there is a transition between S$_\text{c}$ and C confinement in the phase plane of gauge-Higgs theories, and we will also explain what symmetry is actually broken in the Higgs phase of a gauge-Higgs theory.}

\FullConference{XIII Quark Confinement and the Hadron Spectrum - Confinement2018\\
		31 July - 6 August 2018\\
		Maynooth University, Ireland}

\begin{document}

\section{Introduction}

Suppose we have an SU($N$) gauge theory with matter fields in the fundamental representation, e.g. QCD. Wilson loops have perimeter-law falloff asymptotically, Polyakov lines have a non-zero VEV, so what does it mean to say that such theories (QCD in particular) are confining? Most people take it to mean ``color confinement,'' or ``C confinement''  for short, meaning that there are only color neutral particles in the asymptotic spectrum.   The problem with C confinement is that it also holds true for gauge-Higgs theories, deep in the Higgs regime, where there are only Yukawa forces, no linearly rising Regge trajectories,  and no color electric flux tubes. If C confinement is ``confinement,'' then the Higgs phase is also confining.

    We know that the Higgs regime is C confining for several reasons.  First there is the Elitzur theorem \cite{Elitzur:1975im}, which tells us that a local gauge symmetry cannot be spontaneously broken.  Secondly there is the Fradkin-Shenker-Osterwalder-Seiler theorem \cite{Fradkin:1978dv,Osterwalder:1977pc} which proves that there is no transition in coupling-constant space which isolates the Higgs phase from a confinement-like phase.  Finally there is the work of Fr\"olich-Morchio-Strocchi \cite{Frohlich:1980gj}, and also  't Hooft \cite{tHooft:1979yoe}, showing that  physical particles (e.g.\ W's) in the spectrum are created by gauge-invariant operators in the Higgs region. 

    However, in a pure SU($N$) gauge theory there is a different and stronger meaning that can be assigned to the word ``confinement,'' which goes beyond C confinement.   Of course the spectrum consists only of color neutral objects: glueballs.    
 But such theories {\it also} have the property that the static quark potential rises linearly or, equivalently, that large planar
Wilson loops have an area-law falloff.  We may ask:  Is there any way to generalize this property to gauge theories with matter in the fundamental representation?
     
\section{S$_\text{c}$ confinement}
    
    In fact the Wilson area-law criterion for pure gauge theories is equivalent to a property we will call ``separation-of-charge confinement'', or ``S$_\text{c}$ confinement.''  Consider a static $q\overline{q}$ pair, separated by a distance $R$, connected by a Wilson line.  This state evolves in Euclidean time to some lower energy state 
\beq
     \Psi_V \equiv  \overline{q}^a(\vx) V^{ab}(\vx,\vy;A) q^b(\vy) \Psi_0  \ ,
\eeq
where $\Psi_0$ is the ground state, and $V(\vx,\vy;A)$ is a gauge bi-covariant operator transforming under a gauge transformation $g(x)$ as $V^{ab}(\vx,\vy;A) \ra   g^{ac}(\vx,t) V^{cd}(\vx,\vy;A) g^{\dg db}(\vy,t)$.   Let $E_V(R)$ be the energy of this state above the vacuum energy ${\cal E}_{vac}$.   We define S$_\text{c}$ confinement to mean that there exists an asymptotically linear
function $E_0(R) \ra \s R$ at large $R$, such that for {\it any} choice of bicovariant $V$, $E_V(R)\ge E_0(R)$.  For an SU($N$) pure gauge theory, $E_0(R)$ is the ground state energy of a static quark-antiquark pair, and $\s$ is the
string tension.  This is equivalent to the Wilson area-law criterion.

    Our proposal is that S$_\text{c}$ confinement should also be regarded as the confinement criterion in gauge+matter theories.
The crucial element is that the bi-covariant operators $V^{ab}(\vx,\vy;A)$ must depend only on the gauge field $A$ at a fixed time, and not on the matter fields.   The idea is to study the energy $E_V(R)$ of physical states with large separations $R$ of static color charges, unscreened by matter fields. If $V^{ab}(\vx,\vy;A)$ would also depend on the matter field(s), then of course  it is easy to violate the S$_\text{c}$ confinement criterion, e.g.
let $\phi$ be a matter field in the fundamental representation, and let $V^{ab}(\vx,\vy,\phi) = \phi^a(\vx) \phi^{\dg b}(\vy)$. 
Then
\beq
            \Psi_V = \{ \overline{q}^a(\vx) \phi^a(\vx) \} \times \{ \phi^{\dg b}(\vy)q^b(\vy) \}\Psi_0  
\eeq
corresponds to two color singlet (static quark + Higgs) states, only weakly interacting at large separations.  Operators $V$ of this
kind, which depend on the matter fields, are excluded. This also means that the lower bound $E_0(R)$, unlike in pure gauge theories, is not the lowest energy of a state containing a static
quark-antiquark pair. Rather, it is the lowest energy of such states when color screening by matter is excluded.
 
  Consider in particular a unimodular $|\phi|=1$ Higgs field.  In SU(2) the doublet can be mapped to an SU(2) group element  
\bea
            \vec{\phi} =\left[ \begin{array}{c} \phi_1 \cr \phi_2  \end{array} \right] \Longrightarrow
                  \phi = \left[ \begin{array}{rr}
                                               \phi_2^* & ~\phi_1  \cr
                                               -\phi_1^*  & ~\phi_2  \end{array} \right]  \ ,
\eea
and the corresponding action is 
\bea
     S = \beta \sum_{plaq} \oh \mbox{Tr}[UUU^\dg U^\dg] 
       + \gamma \sum_{x,\m} \oh \mbox{Tr}[\phi^\dg(x) U_\m(x) \phi(x+\widehat{\m})]  \ .
\label{S}
\eea
The first question to ask is: Does S$_\text{c}$ confinement exist {\it anywhere} in the $\beta-\g$ phase diagram, apart from pure gauge theory ($\g=0$)?  The answer is yes.  We can show that gauge-Higgs theory is \\
S$_\text{c}$ confining at least for strong couplings,
with $\b \ll 1, \g < 1/10$.  This is based on strong-coupling expansions and the Gershgorim Theorem in linear algebra.  The argument is, however, a little lengthy, and for that we must refer the reader to section VI in our article \cite{Greensite:2018mhh}. 
The second question is whether S$_\text{c}$ confinement
holds {\it everywhere} in the $\beta-\gamma$ phase diagram.  The answer to that is no.   We can construct $V$ operators which violate the S$_\text{c}$ confinement criterion when $\g$ is large enough.  The conclusion is that there must exist a transition between S$_\text{c}$ and C confinement.

    We expand on this second point.  Away from strong coupling, where we can demonstrate its existence, there is no guarantee of S$_\text{c}$ confinement.   But if we can find even one $V$ at some $\beta,\gamma$ such that $E_V(R)$ does not grow linearly with $R$, then S$_\text{c}$ confinement is lost at that $\beta,\gamma$.  For $V=$ a Wilson line, $E_V(R) \propto R$  even for non-confining theories, so this is not a very useful test operator. Instead we consider
\begin{enumerate}
\item {\bf The Dirac state}:  a generalization of the lowest energy state with static charges in an abelian theory. 
\item  {\bf Pseudomatter}:   We introduce fields built from the gauge field which transform like matter fields, and check whether these induce string-breaking.
\item  {\bf  "Fat link" states}:  These are Wilson lines built from links constructed by a smoothing procedure, commonly used as a noise reduction method in lattice gauge theory.
\end{enumerate}

    For any choice of $V$ operator, the energy expectation value is
\beq
            E_V(R) =  -\lim_{t\ra 0} {d\over dt} \log \left[  \langle \Psi_V| e^{-Ht} |\Psi_V \rangle  \right]  - {\cal E}_{vac}  \ .
\eeq
On the lattice, the corresponding quantity is
\beq
        E_V(R) =  - \log \left[ {\left\langle \tr\left[U_0(\vx,t) V(\vx,\vy,t+1) U^\dagger_0(\vy,t) V(\vy,\vx,t) \right] \right\rangle 
               \over \left\langle \tr\left[V(\vx,\vy,t) V(\vy,\vx,t) \right] \right\rangle} \right]  \ .
\eeq

\section{Testing S$_\text{c}$ confinement}

   We begin with the $V$ operator corresponding to the Dirac state.  
In an abelian theory, the gauge-invariant ground state with static $\pm$ electric charges is known exactly:
\beq
    \Psi_{\overline{q} q} = \{ \overline{q}(\vx) G_C^\dg(\vx;A)\} \times \{ G_C(\vy;A)q(\vy) \}\Psi_0  \ ,
\label{abelian}
\eeq
where
\beq
    G_C(\vx;A) = \exp\left[-i \int d^3z ~ A_i(z) \pa_i {1\over 4\pi |\vx-\vz|} \right] \ .
\eeq
$G_C(\vx,A)$ is the gauge transformation which takes the vector potential $A$ to Coulomb gauge.
The obvious non-abelian generalization is  to define  $V^{ab}(\vx,\vy;A) = G_C^{\dg ac}(\vx;A) G_C^{cb}(\vy;A)$, where $G_C^{ab}$
is the gauge transformation taking an $A$ field to Coulomb gauge, and use this to construct the non-abelian Dirac state
\bea
\Psi_V &=&   \overline{q}^a(\vx) G_C^{\dg ac}(\vx;A) G_C^{cb}(\vy;A) q^b(\vy) \Psi_0 \non \\ \non \\
           &=&    \overline{q}^c(\vx) q^c(\vy) \Psi_0 ~~~ \mbox{in Coulomb gauge} \ .
\eea
 We then compute, in Coulomb gauge, 
\bea
    E_V(R) =  -  \log \big\langle {1\over N}\tr[U_0({\bf 0},0) U_0^\dg({\bf R},0)] \big\rangle 
\eea
 by lattice Monte Carlo.

At $\b=2.2, \g=0.84$ there is a sharp thermodynamic crossover.  In Fig.\ 1 we display $E_V(R)$ below ($\g=0.83$), at
($\g=0.84$) and just above ($\g=0.85$) the crossover.
\begin{figure}[t!]
\label{bb22}

\subfigure[~]  
{   
 \includegraphics[scale=0.37]{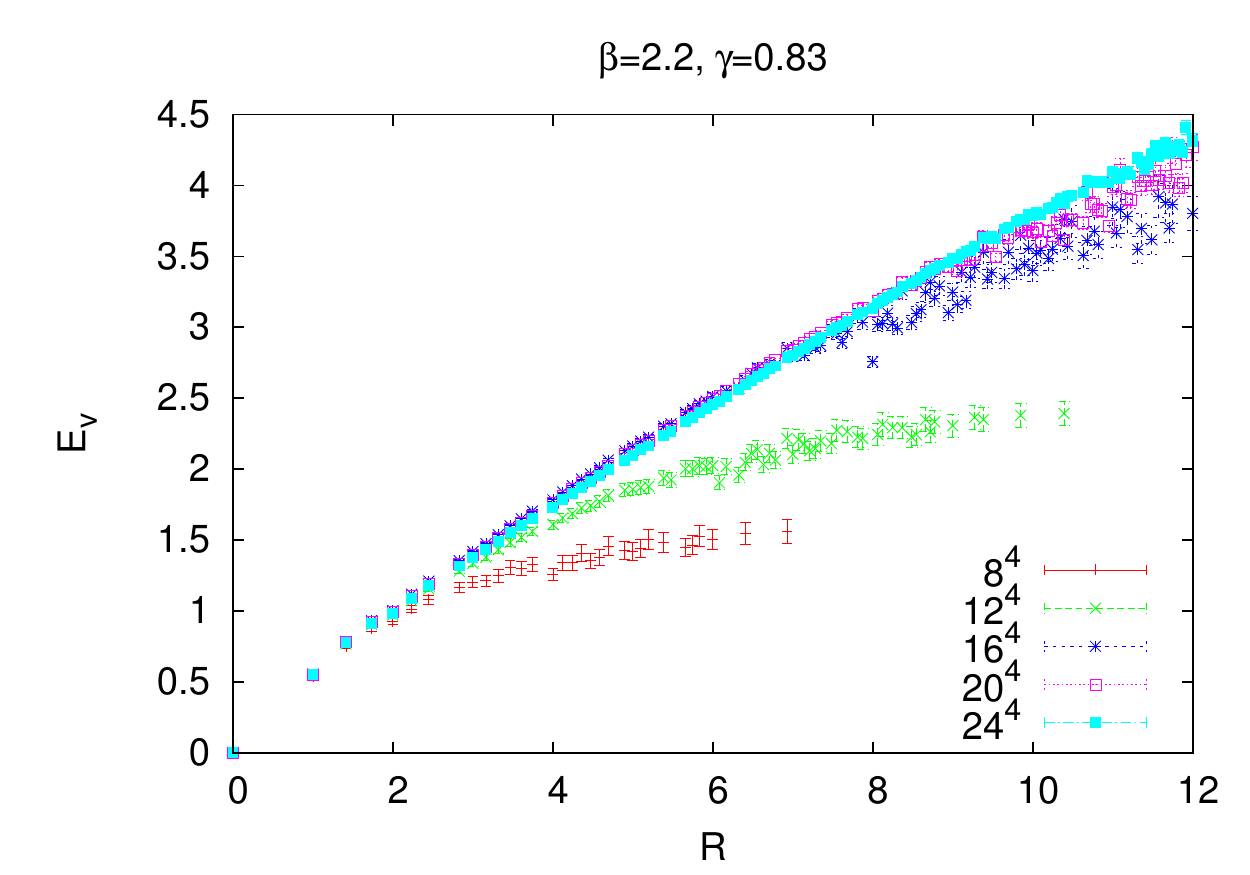}
}
\subfigure[~]
{   
 \includegraphics[scale=0.37]{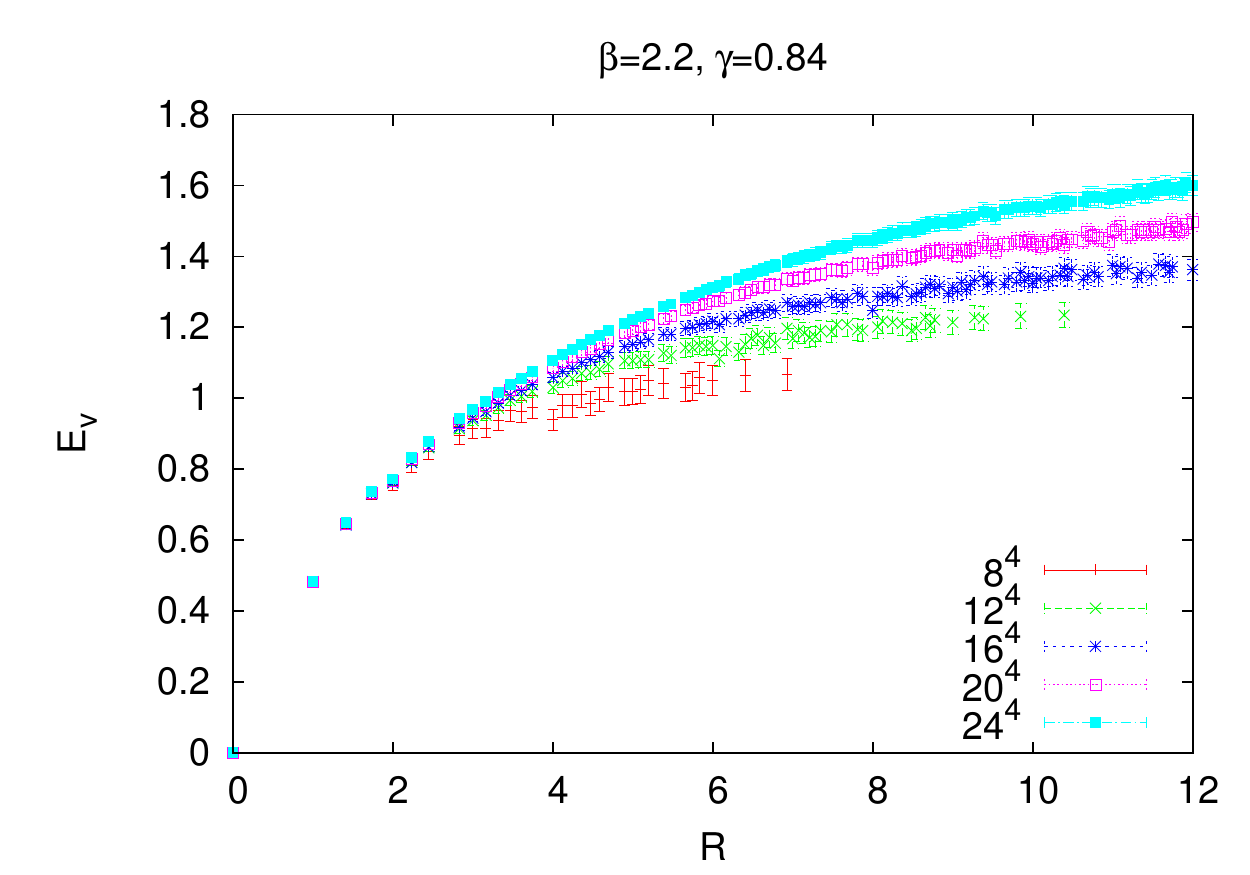}
}
\subfigure[~]
{   
 \includegraphics[scale=0.37]{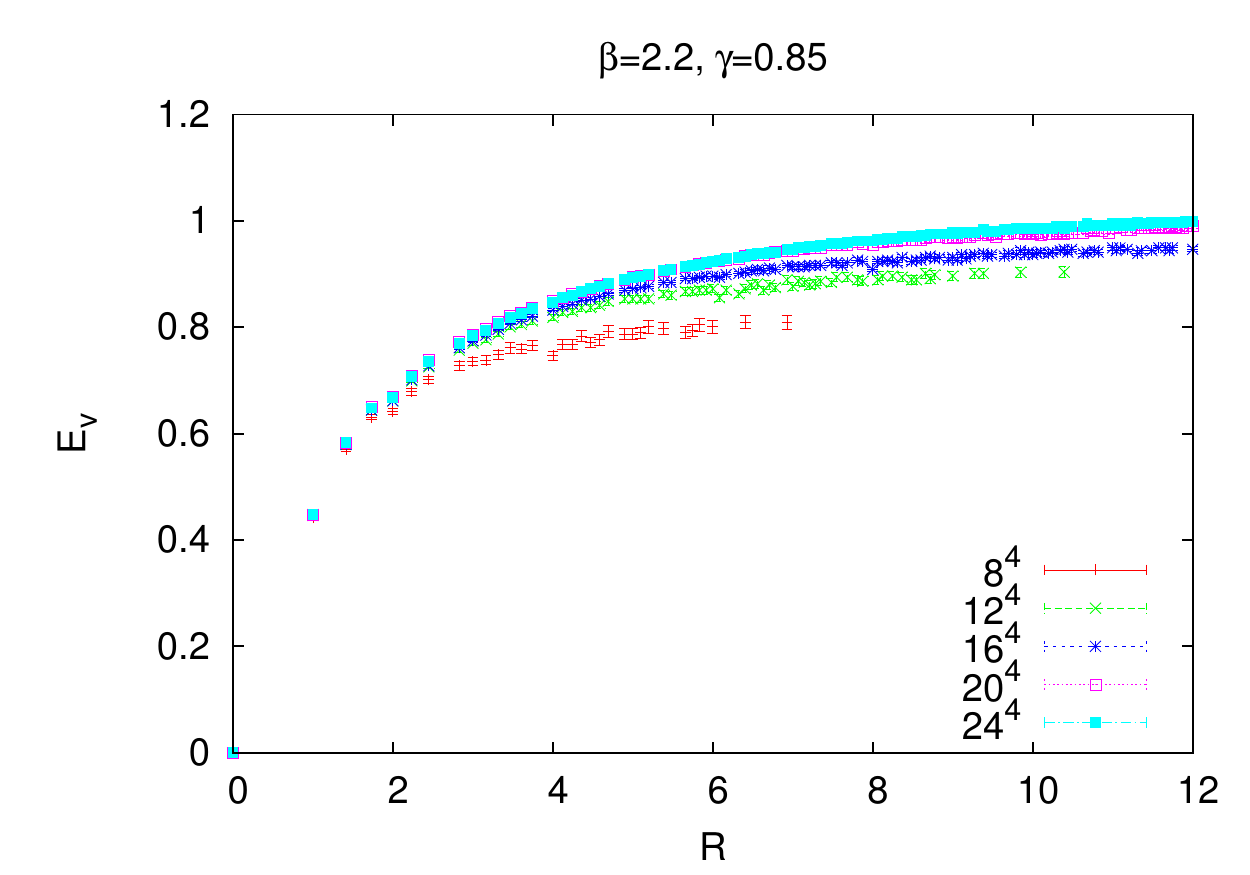}
}
\caption{$E_V(R)$ vs.\ $R$ for the Dirac states in the gauge-Higgs model at $\b=2.2$. (a) just below the thermodynamic crossover at $\g=0.83$; (b) very close to the crossover, at $\g=0.84$; (c) just above the crossover at $\g=0.85$.}
\end{figure}

\begin{figure}[t!]
\subfigure[~]  
{   
 \label{g155}
 \includegraphics[scale=0.37]{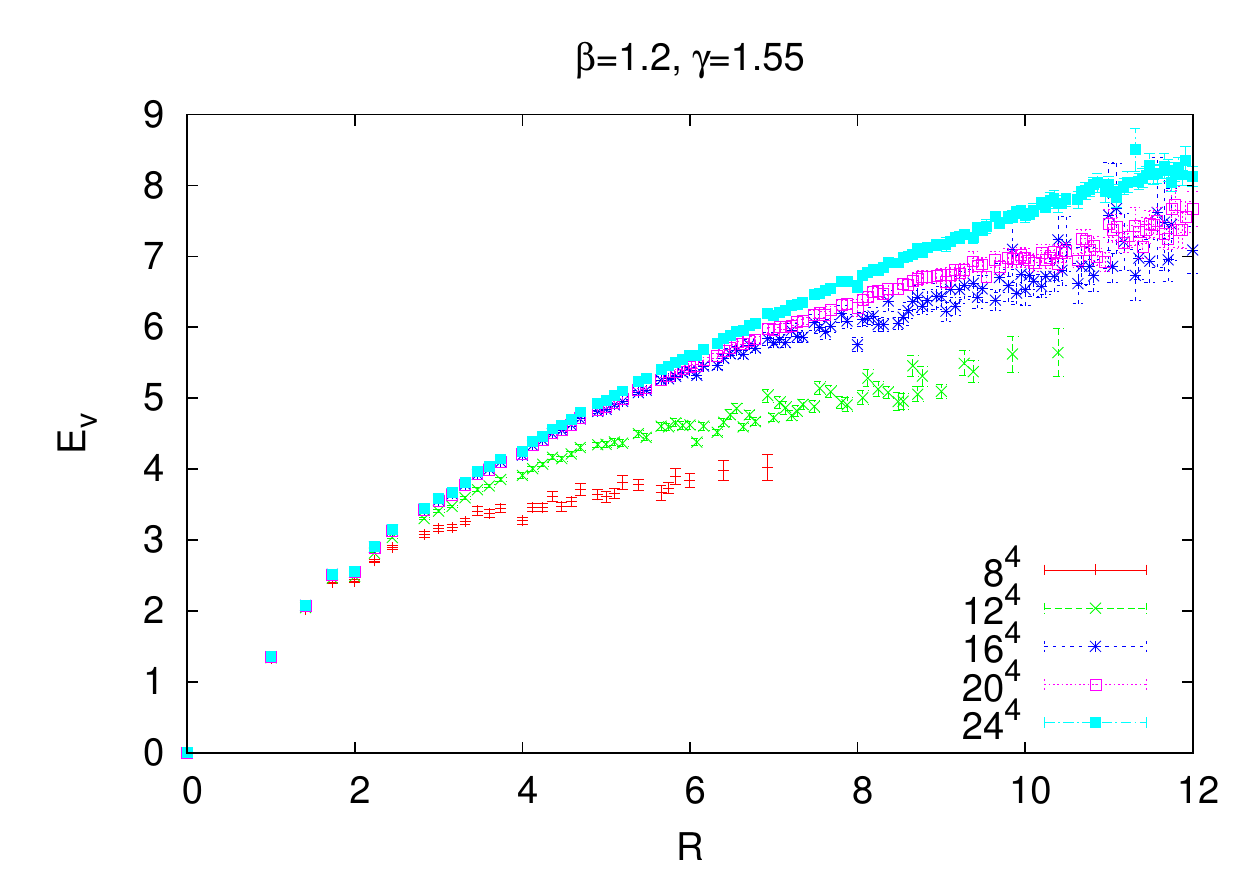}
}
\subfigure[~]
{   
 \label{g168}
 \includegraphics[scale=0.37]{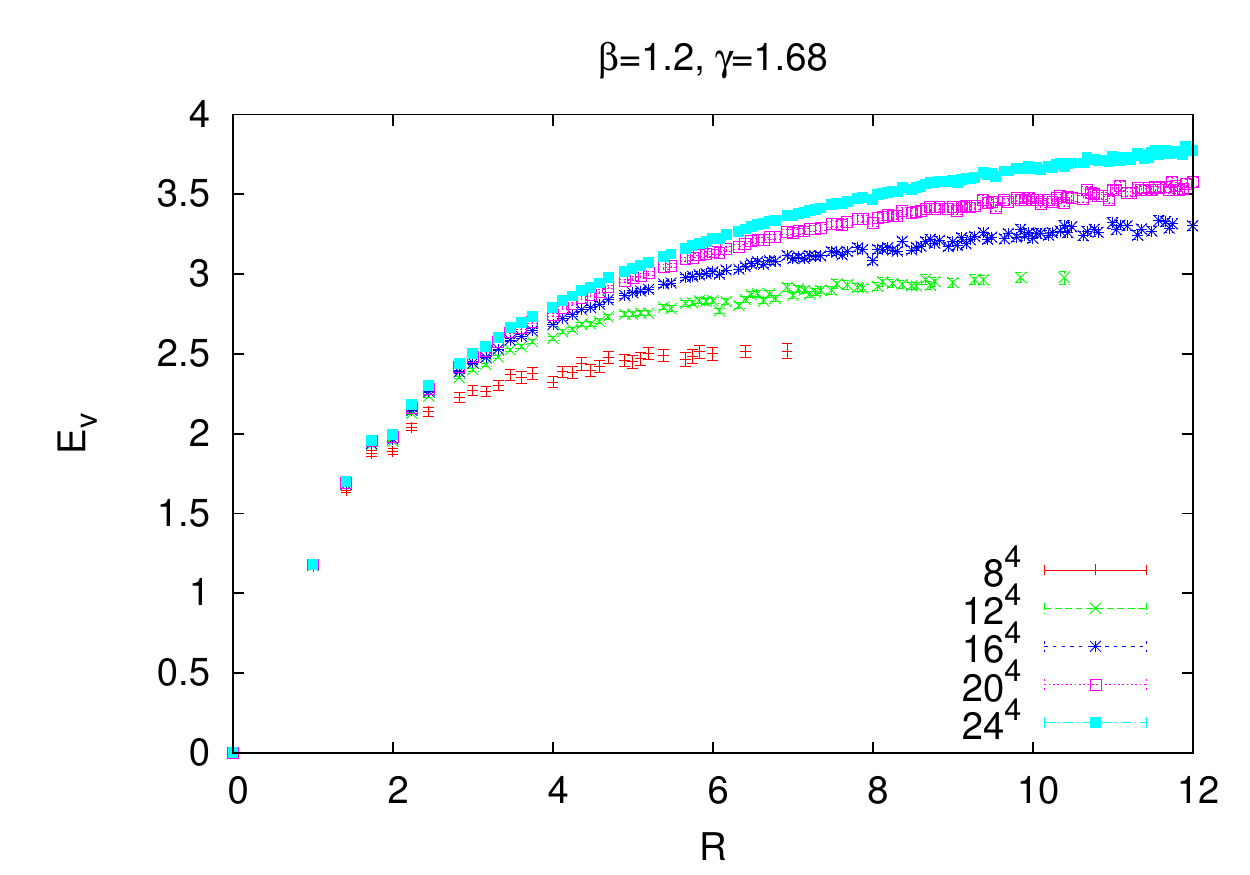}
}
\subfigure[~]
{   
 \label{g1755}
 \includegraphics[scale=0.37]{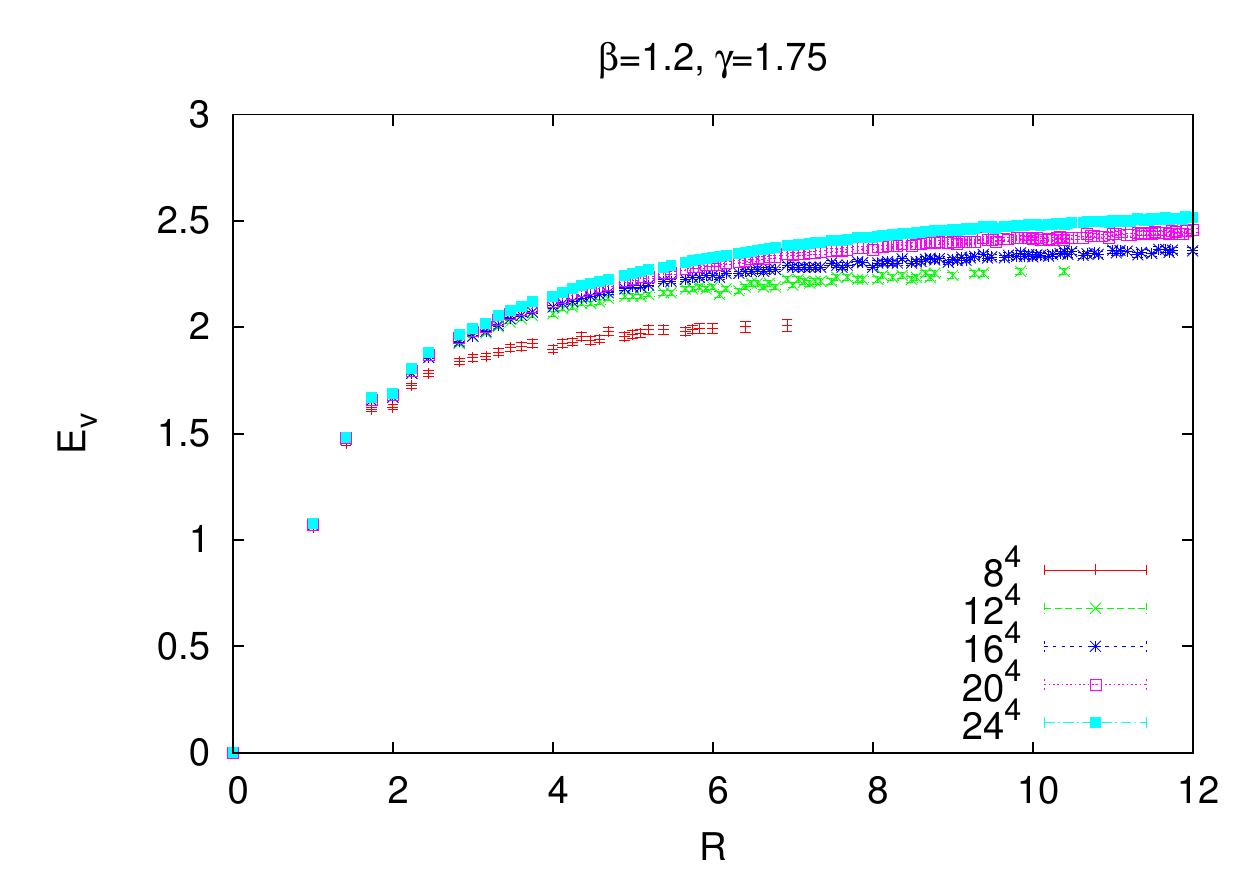}
}
\caption{$E_V(R)$ vs.\ $R$ for the Dirac states in the gauge-Higgs model, as in Fig.\ 1, but this time at $\b=1.2$. (a) $\g=1.55$; (b)  $\g=1.68$; (c) $\g=1.75$.  There is no thermodynamic crossover, but, as in Fig.\ 1, these points are just below, roughly at, and just above a Coulomb gauge remnant symmetry breaking transition \cite{Caudy:2007sf}.}
\label{b12}
\end{figure}
$E_V(R)$ rises linearly below the crossover in the large volume limit,  consistent with (but not a proof of) S$_\text{c}$ confinement in this region.  Above the crossover, where $E_V(R)$ levels out, the system is definitely in a C confining regime. 
Note that if, for some $V$,  $E_V(R)$ is bounded from below by a linear potential, then this is a necessary but not sufficient condition for  S$_\text{c}$ confinement; the bound must hold for all $V$.  If $E_V(R)$ violates that bound, then this is a sufficient but not necessary condition for C confinement, which holds if there exists even one $V$ that violates the bound.

    The transition in the Dirac state between S$_\text{c}$ and C behavior coincides with the thermodynamic crossover, where there is such a crossover, but persists into the strong-coupling region where the crossover is absent.  In Fig.\ \ref{b12} we show the
corresponding behavior at $\b=1.2$, where we find a transition from S$_\text{c}$ to C behavior at about $\g=1.68$.  Again,
this means that the region $\g>1.68$ is C confining, but the S$_\text{c}$ behavior at $\g<1.68$ is a necessary but not sufficient condition that the region is S$_\text{c}$ confining.

   The  transition from S$_\text{c}$ to C behavior seen in $E_V(R)$ for the Dirac state coincides with the breaking of a remnant gauge symmetry $g(x,t)=g(t)$ that exists in Coulomb gauge. The appropriate order parameter for the symmetry breaking on a time slice is
\beq
              u(t) =  {1\over \sqrt{2} V_3} \sum_{\vx} U_0(\vx,t) \ .
\eeq
Other gauges, however, have other remnant symmetries, and it was found in \cite{Caudy:2007sf} that the transition lines for remnant-symmetry breaking are gauge-dependent.

    A second test is to compute $E_V(R)$ corresponding to pseudomatter states, with $V$ built from pseudomatter fields.  A pseudomatter field is a field constructed from the gauge field which transforms like matter in the fundamental representation.
An example is any eigenstate 
\beq
           (-D_i D_i)^{ab}_{\vx \vy} \vph^b_n(\vy) = \lambda_n \vph^a_n(\vx) 
\eeq
of the covariant spatial Laplacian 
\bea
 (-D_i D_i)^{ab}_{\vx \vy} =  \sum_{k=1}^3 \left[2 \delta^{ab} \delta_{\vx \vy} - U_k^{ab}(\vx) \delta_{\vy,\vx+\hat{k}} 
       - U_k^{\dg ab}(\vx-\hat{k}) \delta_{\vy,\vx-\hat{k}}  \right]    \ .
\eea
We construct
\beq
           V^{ab}(\vx,\vy;A) = \vph_1^a(\vx) \vph_1^{\dg b}(\vy) 
\eeq
from the lowest-lying eigenstate, and compute $E_V(R)$ by lattice Monte Carlo.

    Our third $V$ operator is a Wilson line running between the static quark-antiquark charges, built from ``fat links.''  These are constructed by an iterative procedure. Let $U^{(0)}_k(x)=U_k(x)$, and define 
 \bea
 U^{(n+1)}_i(x) 
 &=& \N \bigg\{ \a U^{(n)}_i(x) + \sum_{j\ne i} \left( U^{(n)}_j(x) U^{(n)}_i(x+\hat{j}) U^{(n)\dg}_j(x+\hat{i})   \right.  \non \\
 & & \qquad \qquad \qquad  \left. + U^{(n)\dg}_j(x-\hat{j}) U^{(n)}_i(x-\hat{j}) U^{(n)}_j(x-\hat{j}+\hat{i}) \right)   \bigg\}  \ ,
\eea
where $\a$ is a constant, $i,j \ne 4$, and $\N$ is an overall constant such that $U_i^{(n+1)}$ is an SU(2) group element. Denote the link variables after the last iteration as $U^{fat}_i(x)$ and define, for $y=x + R \hat{k}$,
 \bea
       V_{fat}(x,y;A) &=& U^{fat}_k(x)U^{fat}_k(x+\hat{k})...U^{fat}_k(x+(R-1)\hat{k}) \ .
\eea
We then compute $E_V(R)$ for $V=V_{fat}$.

   Numerical results for $E_V(R)$ corresponding to pseudomatter and fat link states are found in \cite{Greensite:2017ajx}, but briefly the results are
\begin{itemize}
\item We find an S$_\text{c}$-to-C confinement transition for the $V$ operator constructed from pseudomatter fields.  The transition line is close to (but a little below) the transition line for the Dirac state.
\item The fat link state seems to be everywhere S$_\text{c}$ confining.  This doesn't mean that the gauge-Higgs theory is everywhere S$_\text{c}$ confining.  It means instead that not every operator can
detect the transition to C confinement. 
\end{itemize}
    
\section{Comments on other criteria}
    
Other criteria for distinguishing the confinement from the Higgs phase have been proposed in the past, in particular  (i) the Kugo-Ojima criterion \cite{Kugo:1979gm}; (ii) Non-positivity/unphysical poles in quark/gluon propagators, see e.g.\ \cite{Stingl:1985hx}; and (iii) the Fredenhagen-Marcu criterion  \cite{Fredenhagen:1985ft}. The first two of these criteria assume the existence of BRST symmetry, which is problematic non-perturbatively for the following reasons:
\begin{enumerate}
\item The Neuberger 0/0 problem \cite{Neuberger:1986xz}. BRST symmetry implies the vanishing of the functional integral in covariant gauges.
\item BRST symmetry is broken by the gauge fixing procedure employed in lattice Monte Carlo simulations, which restricts the domain of configurations to the Gribov region inside the first Gribov horizon \cite{Cucchieri:2014via}. 
 \item BRST perturbative analysis yields the wrong spectrum for the SU(3) gauge-Higgs model, even deep in the
 Higgs region \cite{Maas:2018xxu}.
\end{enumerate}
As for the Fredenhagen-Marcu criterion, this does not really distinguish a Higgs from a confinement phase.  Rather, it is designed to distinguish between a massive phase and a free charge (or ``Coulomb'') phase.  It fails to even distinguish between 
the confined phase and the Higgs phase in an SU(2) gauge theory with matter in the adjoint representation \cite{Azcoiti:1987ua}, where there is a clear thermodynamic separation of the confining and Higgs phases, and a spontaneous breaking of global $Z_2$ symmetry
in the Higgs phase.

\section{The Brout-Englert-Higgs mechanism and symmetry breaking}

    Does the transition from S$_\text{c}$ to C confinement correspond to the spontaneous breaking of some symmetry in the gauge-Higgs theory?

     The obvious answer is no.  A local gauge symmetry cannot break spontaneously, as we know from Elitzur's theorem.
And the breaking of some other symmetry, a global continuous symmetry of some kind, would necessarily be accompanied
by massless Goldstone particles, which are absent in the theory.  These two facts would seem to conclusively rule out understanding
the BEH mechanism, and the distinction between S$_\text{c}$ and C confinement, in terms of symmetry breaking.  But let us look anyway at the global symmetry of the SU(2) gauge-Higgs model with action \rf{S}.
It is well known, in the SU(2) gauge-Higgs model, that the full symmetry of the Higgs action is 
SU(2)${}_{gauge} \times$ SU(2)${}_{global}$:  
 \bea  
             U_\m(x) &\ra& L(x) U_\m(x) L^\dg(x+\hat{\mu}) \non \\
             \phi(x) &\ra& L(x) \phi(x) R \ ,
\eea
where $L(x)$ is a local, and $R$ is a global SU(2) transformation.  SU(2)$_{gauge}$ can't break spontaneously, but what about SU(2)${}_{global}$?  Note that $Z$, the partition function, is a sum of ``spin systems''
\beq
            Z = \int DU ~ Z_{spin}(U) e^{-S_W[U]}  \ ,
\eeq
where
\bea
          Z_{spin}(U) &=& \int D \phi ~ e^{-S_H[\phi,U]} \non \\
                                     &=& e^{-{\cal F}_H[\g,U]} \ ,
\eea 
and where $S_W,S_H$ are the Wilson and Higgs components, respectively, of the SU(2) gauge-Higgs action \rf{S}.  The only symmetry of the spin system, since $U_\m(x)$ is fixed, is the SU(2)${}_{global}$ symmetry $\phi(x) \ra \phi(x) R$.  
 It is possible that the SU(2)${}_{global}$ ($R$-transformation) symmetry breaks in {\it each} $Z_{spin}(\g,U)$ without breaking in the {\it sum} over spin systems.   This might be a gauge-invariant version of the gauge-dependent statement that $\langle \phi \rangle \ne 0$, and possible way to evade the Goldstone modes.
 
    We can construct a gauge-invariant order parameter to detect the symmetry breaking in $Z_{spin}$.   Consider $\phi(x)$ fluctuating in a background gauge field $U$, which is held fixed.  Denote its average
value in this background as $\overline{\phi}(x;U)$, i.e.
\bea
          \overline{\phi}(x;U) &=& {1\over Z_{spin}[U]} \int D\phi ~ \phi(x) \exp\left[\g \sum_{x,\m} \oh \tr[\phi^\dg(x) U_\m(x) \phi(x+\hat{\m})]\right]   \non \\
          Z_{spin}[U] &=& \int D\phi  ~ \exp\left[\g \sum_{x,\m} \oh\tr[\phi^\dg(x) U_\m(x) \phi(x+\hat{\m})]\right] \ .
\eea
In general, $\int dx \overline{\phi}(x;U) = 0$, because if no gauge is fixed, so $U_\m(x)$ varies wildly in space, then $\phi(x)$ also varies wildly.   On the other hand, it could be that $\overline{\phi}(x;U) \ne 0$
at any given point $x$, even if the spatial average vanishes.  Since the action at fixed $U_\m$ is invariant under $\phi(x) \ra \phi(x) R$,  this would imply the spontaneous symmetry breaking of an SU(2)${}_{global}$ symmetry in $Z_{spin}(\g,U)$, while $\langle \overline{\phi}(x,U) \rangle = 0$, as it must, in the full theory.

   We therefore introduce and compute the following gauge-invariant order parameter: 
\bea
         \langle \Phi \rangle&=& \left\langle \sqrt{\oh \tr[\overline{\phi}^\dg(x;U)\overline{\phi}(x;U)]} \right\rangle  
              = {1\over Z} \int DU D\phi \sqrt{\oh \tr[\overline{\phi}^\dg(x;U)\overline{\phi}(x;U)]} e^{S[U,\phi]} \ ,
\eea
which is evaluated by a Monte Carlo-within-a-Monte Carlo procedure.   That is to say, the usual update sweeps involve
sweeping site by site through the lattice, and updating the four
link variables and the Higgs field at each site. Since both the
link and scalar field variables are elements of the SU(2) group,
the updates of both types of variables can be carried out using
the Creutz heat bath method. The data-taking sweep, however, is a simulation of the spin
system $Z_{spin}[U]$, and entails $n_{sw}$ sweeps through the lattice, updating
only the Higgs field by the heat bath method, while keeping
the gauge field fixed.   In the course of this data-taking sweep, on a
finite lattice volume $V$, we measure
\beq
           \Phi_{n_{sw},V}[U] = {1\over V} \sum_x \left|{1\over n_{sw}} \sum_{t_5=1}^{n_{sw}} \phi(x,t_5) \right| \ ,
\label{bigphi}
\eeq
where $\phi(x,t_5)$ is the Higgs field at point $x$ after $t_5$ update sweeps, holding the $U$ field fixed.  The quantity we 
would like to estimate is the limiting value
\beq
     \langle \Phi \rangle = \lim_{n_{sw}\ra \infty} \lim_{V\ra \infty} \langle \Phi_{n_{sw},V}[U] \rangle \ ,
\eeq
with the order of limits as shown.  In the infinite volume limit we expect, on general statistical grounds, that
\beq
          \langle \Phi_{n_{sw},\infty}[U] \rangle \approx \langle \Phi \rangle + {\mbox{const.} \over \sqrt{n_{sw}}} \ .
\label{extrapolate}
\eeq
In the unbroken phase, with $\langle \Phi \rangle=0$, this behavior would also hold at finite volume.   But of course there is no spontaneous symmetry breaking on a finite lattice; any ``broken''
state is only metastable in time (just like a real magnet).  ``Time'' in our case is the number of Monte Carlo sweeps $n_{sw}$ used to compute $\overline{\phi}(x;U)$.   In the broken phase,
we therefore expect $\langle \Phi_{n_{sw},V}[U] \rangle \approx \langle \Phi_{n_{sw},\infty}[U] \rangle$ to only hold for $n_{sw}$ smaller than
the lifetime $T_{meta}(V)$ of the metastable state, and then to go to zero as $n_{sw}$ increases beyond $T_{meta}(V)$.
So on a finite volume we must use \rf{extrapolate} to extrapolate, from a set of values $\{\langle \Phi_{n_{sw},V}[U] \rangle\}$ computed at $n_{sw} < T_{meta}(V)$,  to the $n_{sw} \ra \infty$ limit,
checking that $T_{meta}(V)$, where the linear extrapolation breaks down, increases with lattice volume $V$, and that the extrapolated estimate for $\langle \Phi \rangle$ converges as $V$ increases.

\begin{figure}[htb]
\subfigure[~below the transition, $\g=1.2$]  
{   
 \label{fig2a}
 \includegraphics[scale=0.55]{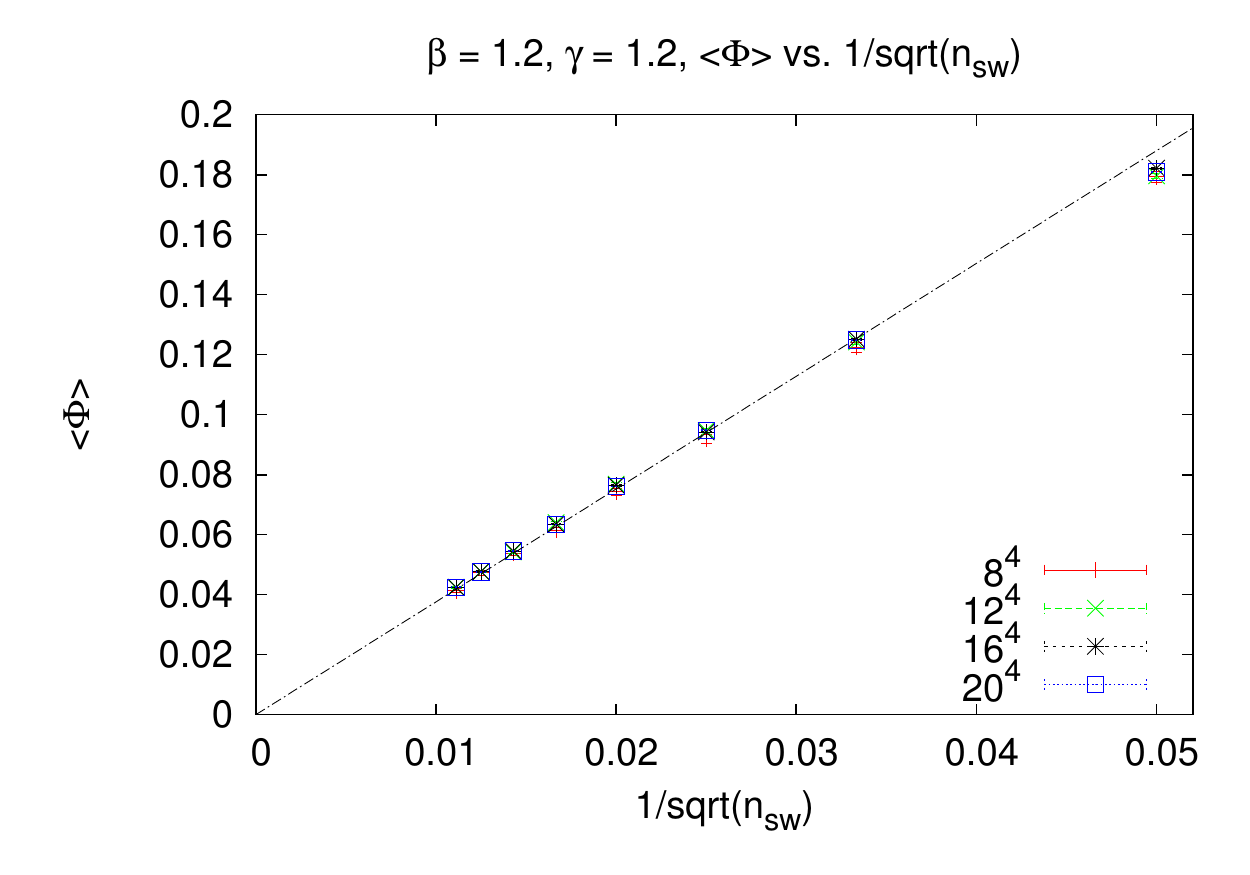}
}
\subfigure[~above the transition, $\g=1.35$]  
{   
 \label{fig2b}
 \includegraphics[scale=0.55]{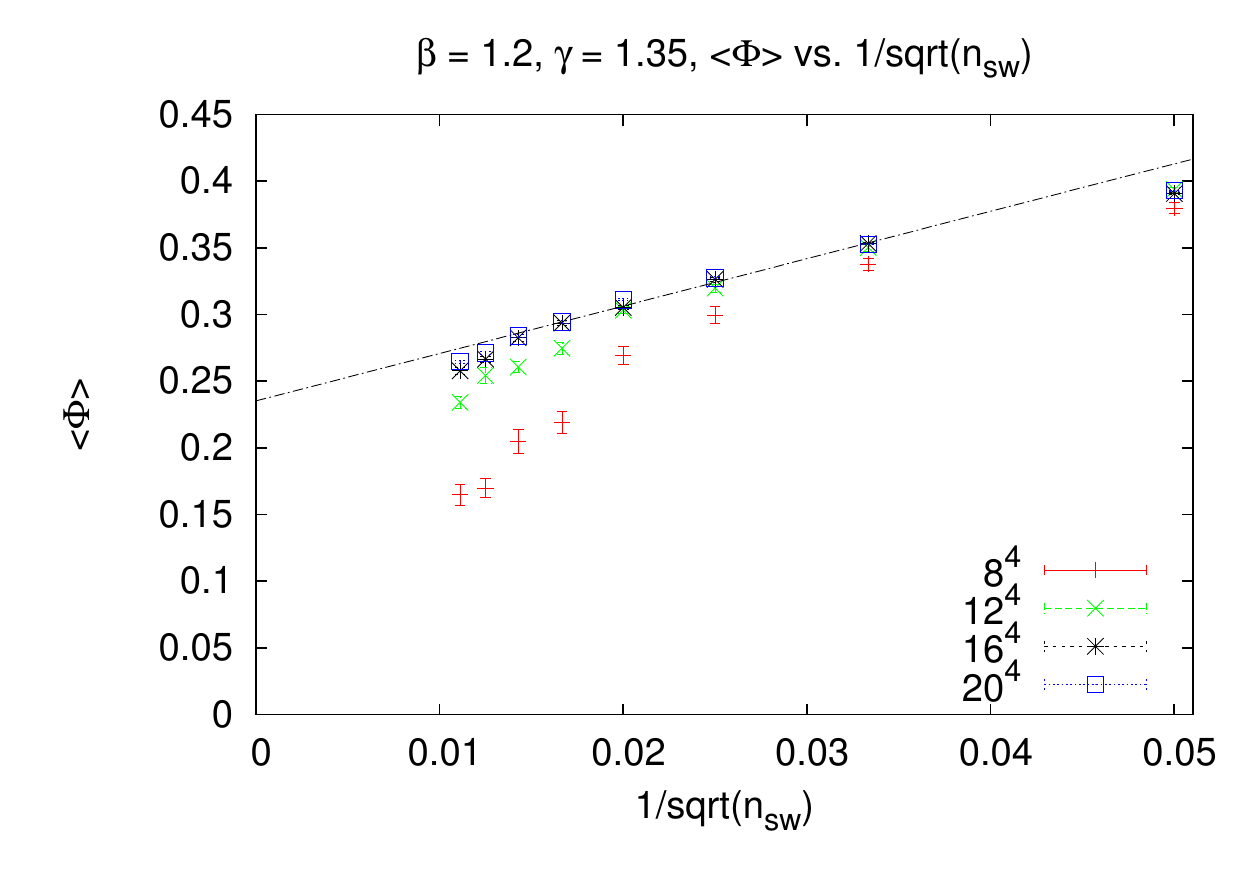}
}
\caption{Gauge invariant order parameter $\langle \Phi \rangle$ vs.\ $1/\sqrt{n_{sw}}$, where
$n_{sw}$ are the number of sweeps carried out on the matter field at fixed gauge field.
The data is for $\b=1.2$ at lattice volumes $8^4,12^4,16^4,20^4$.  (a) below
the transition, at $\g=1.2$; (b) above the transition, at $\g=1.35$. Note the convergence, in subfigure (b),
to a straight line with non-zero intercept on the $y$-axis, as lattice volume increases.}
\label{fig2}
\end{figure}

\begin{figure}[htb]
\centerline{\includegraphics[scale=0.7]{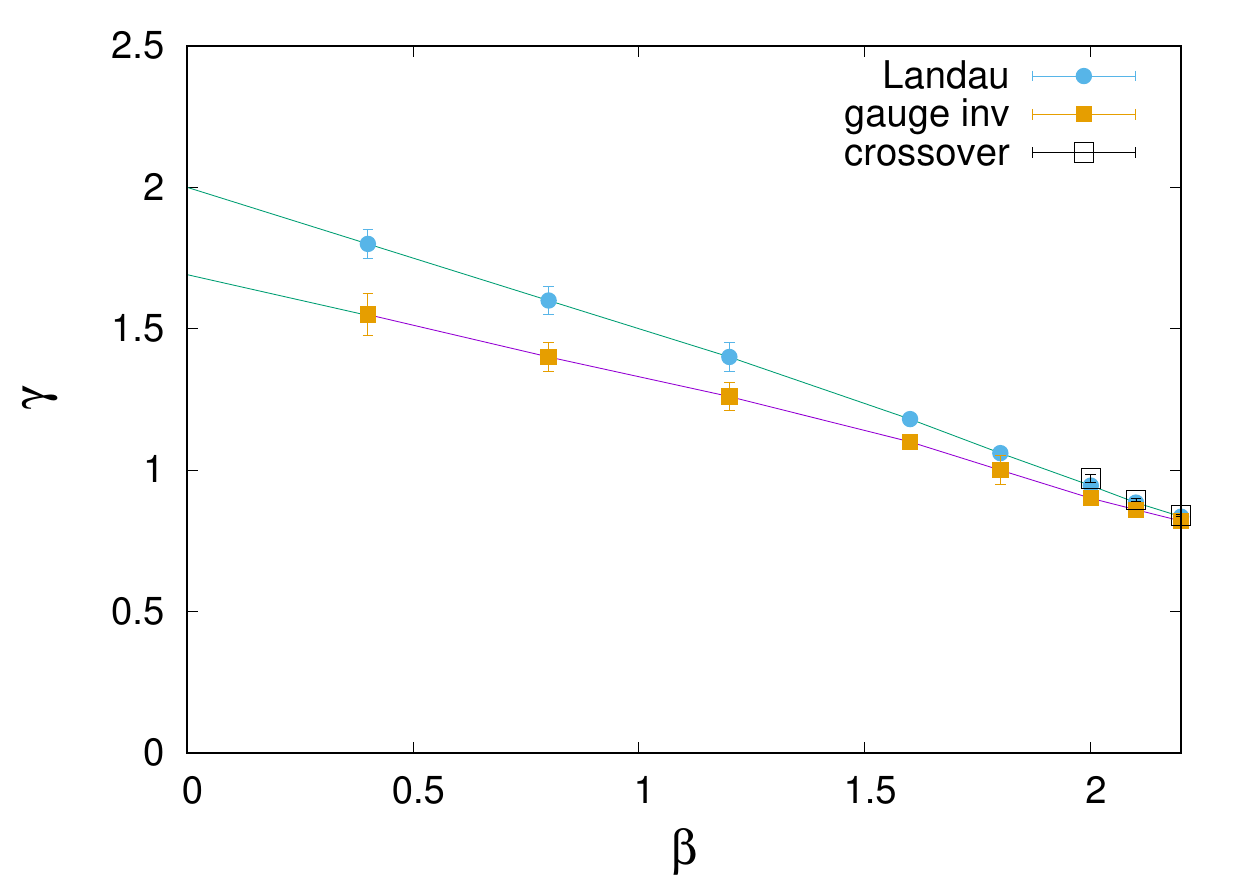}}
\caption{Transition line (square points) for the gauge-invariant global SU(2) symmetry described in the text.  The transition line
for remnant gauge symmetry breaking in Landau gauge (circles) is shown for comparison, along with points at $\b \ge 2.0$ (open squares) where we find a sharp thermodynamic crossover.} 
\label{phase}
\end{figure}

   This behavior is illustrated in Fig.\ \ref{fig2}, at $\b=1.2$, where below the transition $\langle \Phi \rangle$ extrapolates to zero as $n_{sw}\ra \infty$ (Fig.\ \ref{fig2a}), while above the transition $\langle \Phi \rangle$ appears to extrapolate to a non-zero value in the infinite volume limit (Fig.\ \ref{fig2b}), although at any fixed volume it appears to eventually drop to zero.  The transition point, where $\langle \Phi \rangle$ begins to move away from zero, coincides with a peak in an appropriately defined susceptibility.  In this way we can map out the symmetry breaking transition line throughout the phase diagram, which is shown in Fig.\   \ref{phase}.

\subsection{Absence of Goldstone modes}

      The order parameter for symmetry breaking in a $Z_{spin}(\g,U)$ system
 is the gauge covariant quantity $\overline{\phi}(x;U)$, which vanishes when averaged over gauge-field configurations, i.e.\
$ \langle  \overline{\phi}(x;U)  \rangle = 0$.  The symmetry is therefore broken in the Higgs phase in each of the $Z_{spin}[U]$ subsystems, but it is not broken in the full theory.  This is the underlying reason for the absence of physical Goldstone modes: they are gauge variant, and average to zero in the full theory.  
The same can be said of long-range
correlations in various $n$-point functions.  Such long-range correlations only exist, in a theory at fixed $U$ and $\Phi[U] > 0$, in the $n$-point functions of gauge non-invariant operators.  These correlators vanish in the full theory.  To pick a trivial example, the correlator
\beq
             \oh \overline{\tr[\phi^\dg(x) \phi(y)]} = {1\over Z_{spin}(U)} \int D\phi ~ \oh \tr[\phi^\dg(x) \phi(y)] e^{-S_H[\phi,U]}
\eeq
may have long range correlations for a particular gauge field $U$ with $\Phi[U] > 0$, but this quantity vanishes when integrating over all gauge fields,
\beq
\langle \overline{\tr[\phi^\dg(x) \phi(y)]} \rangle = 0 \ ,
\eeq
as does $\langle \tr[\phi^\dg(x) \phi(y)] \rangle$.  
One could, of course, construct a gauge-invariant quantity such as
\beq
G(x,y) =\langle \overline{\tr[\phi^\dg(x) U(x,y)\phi(y)]} \rangle \ ,
\eeq
where $U(x,y)$ is a Wilson line with endpoints $x,y$,
but there is no particular reason why this quantity should have a power-law falloff.  The point here is that long-range correlations in
the individual $Z_{spin}(U)$, which are due
to the Goldstone theorem, must cancel out in the full theory.  

\subsection{Symmetry breaking in SU(3) gauge-Higgs theory}

\begin{figure}[htb]
\centerline{\includegraphics[scale=0.7]{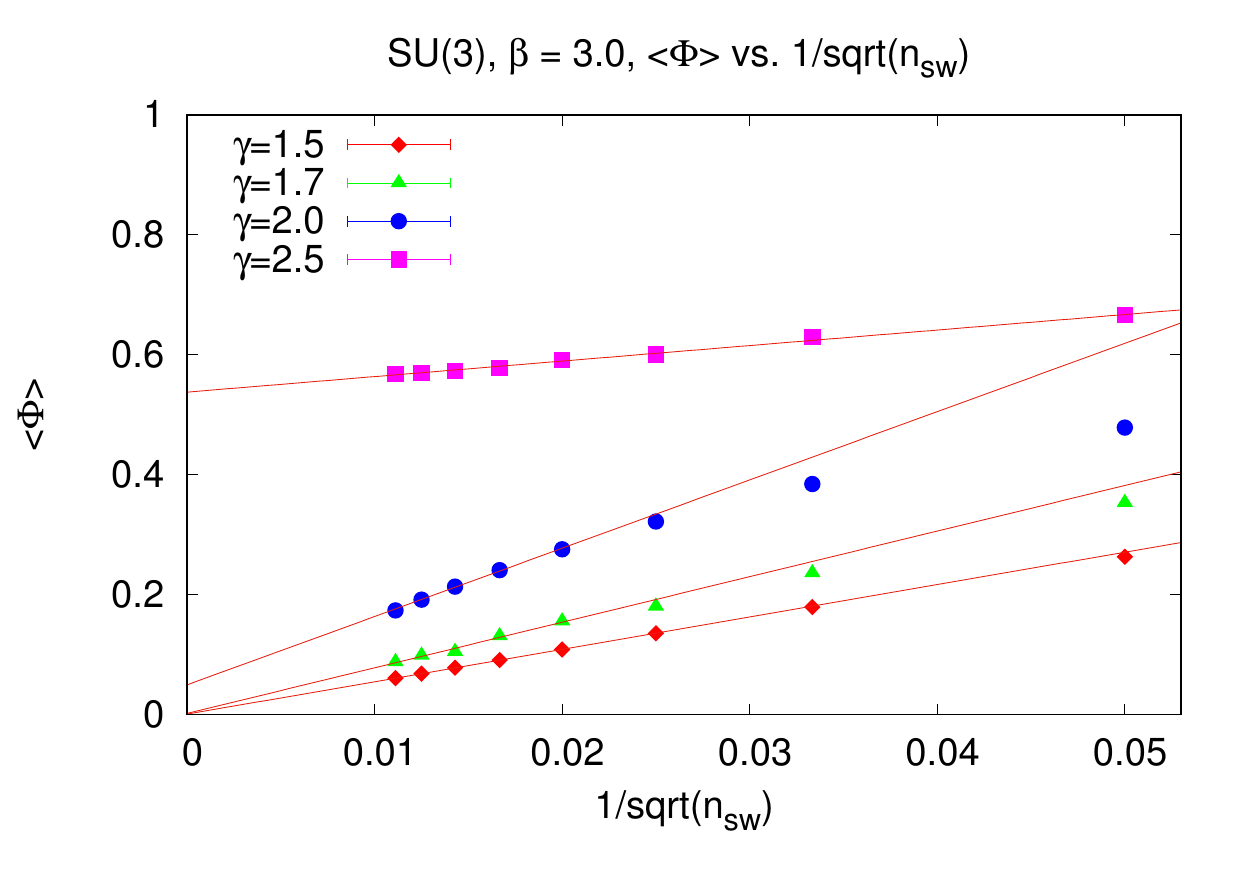}}
\caption{Gauge invariant SU(3) order parameter $\Phi$ vs.\ $1/\sqrt{n_{sw}}$ at $\b=3.0$ on a $16^4$ lattice volume.  Below the transition at $\g=1.85$, the data extrapolates to zero as $n_{sw}\ra \infty$.  Above the
transition, the data extrapolates to non-zero values.} 
\label{opfig}
\end{figure}

\begin{figure}[htb]
\centerline{ \includegraphics[scale=0.7]{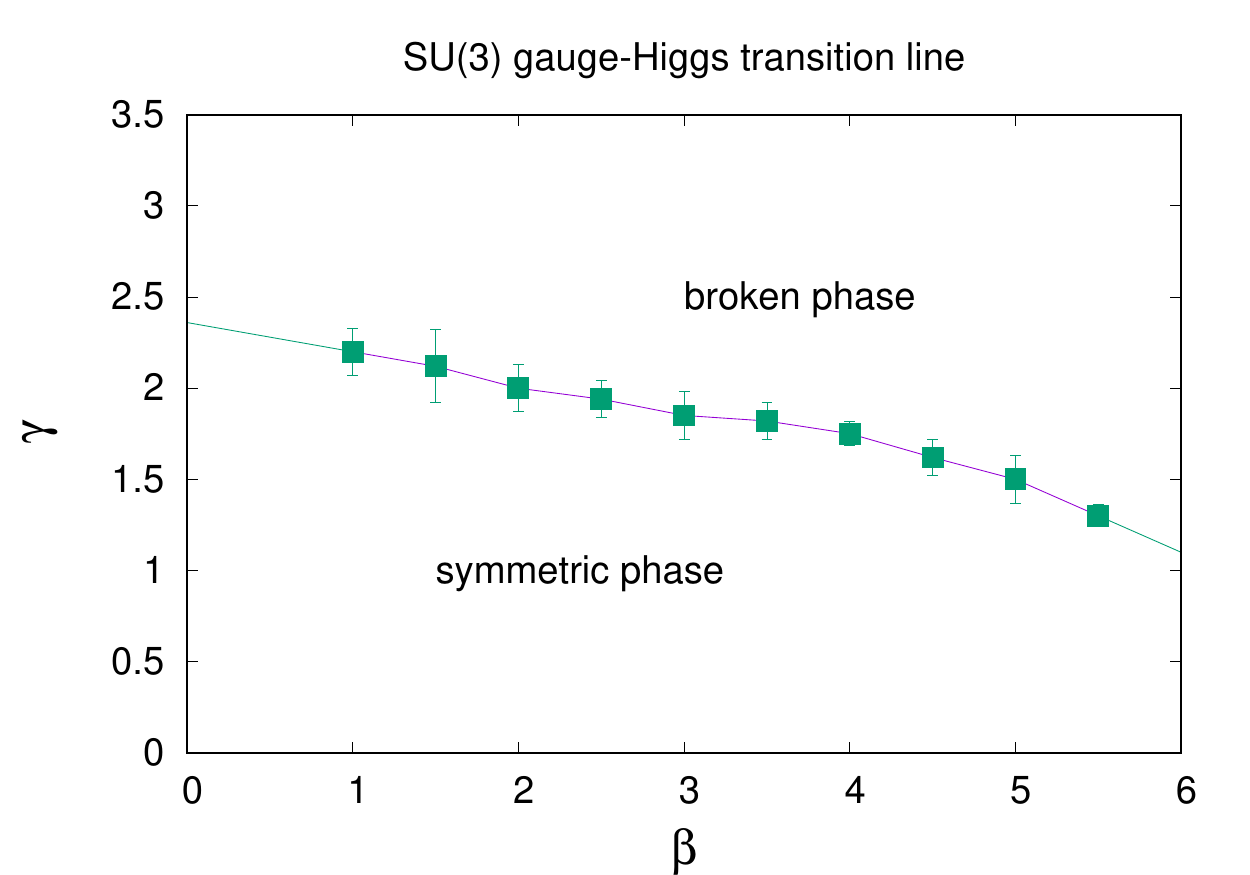}}
\caption{Gauge-invariant transition line for global U(1) symmetry breaking in SU(3) gauge-Higgs theory.} 
\label{su3gb}
\end{figure}

The global ``R'' symmetry in the SU(2) gauge-Higgs model is accidental.  A Higgs field in SU($N$) gauge-Higgs theory at $N>2$ cannot be expressed as an SU($N$) group element.  However, the SU($N>2$) Higgs action
\beq
   S_H[U,\phi] = \gamma \sum_{x,\m} \mbox{Re}[\phi^\dg(x) U_\m(x) \phi(x+\widehat{\m})]  
\label{Shiggs2}
\eeq
does have a global U(1) symmetry, distinct from the gauge symmetry \cite{Maas:2016ngo}:
\beq
           \phi(x) \ra e^{i\theta} \phi(x)  \ ,
\eeq
and this global symmetry can be spontaneously broken  in the same sense as in the SU(2) case.  The order parameter is the same as before, changing only the definition of the gauge invariant modulus
\beq
|\overline{\phi}(x;U)| =\sqrt{\overline{\phi}^\dg(x;U)\overline{\phi}(x;U)}   \ ,
\label{mod2}
\eeq    
where a dot product of color indices, rather than a trace, is implied.  We show in Fig.\ \ref{opfig}, at $\b=3.0$, that 
$\langle \Phi \rangle \ra 0$ as $n_{sw} \ra \infty$ below the
the transition $\g=1.85$, while $\langle \Phi \rangle$ extrapolates to a non-zero value above the transition.
The transition line in the $\b-\g$ coupling plane, for $0<\b<5.6$ is shown in Fig.\ \ref{su3gb}.

\subsection{Symmetry Breaking and the S$_\text{c}$-to-C transition}

\begin{figure}[t!]
\centerline{\includegraphics[scale=0.7]{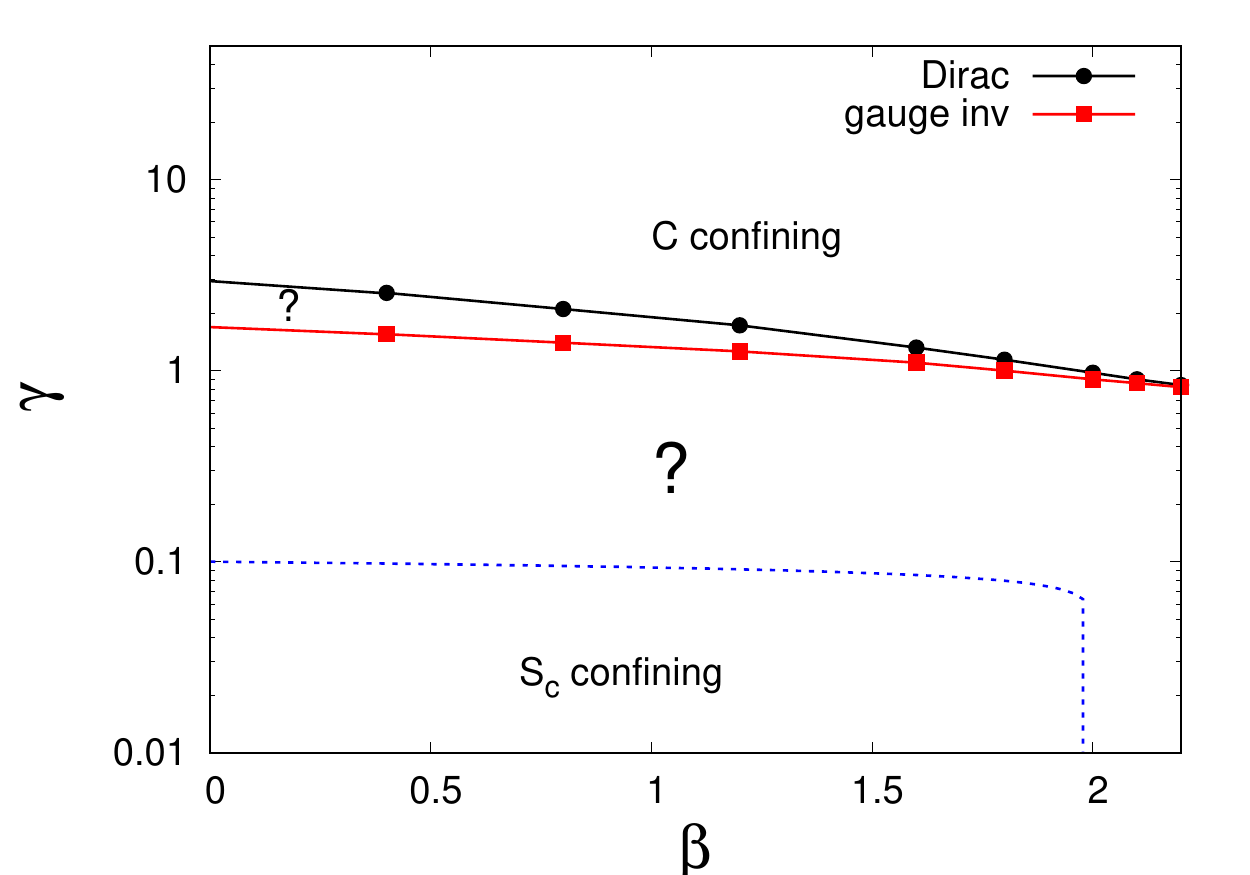}}
\caption{For SU(2) gauge-Higgs theory, C confinement exists above the line denoted ``Dirac,'' and S$_\text{c}$ confinement exists in a 
strong-coupling region, as well as
along the line at $\b=0$.  The location of C and S$_\text{c}$ confinement in the rest of the phase diagram is uncertain.  Our conjecture
is that the S$_\text{c}$-to-C confinement transition line coincides with the gauge-invariant symmetry breaking line, denoted ``gauge inv''
in the figure.  Error bars on data points are, on this scale, smaller than the symbol size. } 
\label{sc}
\end{figure}

    We conjecture that the transition from S$_\text{c}$ to C confinement coincides with the gauge-invariant symmetry breaking transitions seen in Figs.\ \ref{phase} and \ref{su3gb}.    The situation at the moment is illustrated in Fig.\ \ref{sc}, for SU(2) gauge-Higgs theory.  C confinement is known to exist above the Dirac line shown, but we do not know
how far it extends below that line.  S$_\text{c}$ confinement exists inside a strong-coupling region, whose boundary is indicated somewhat schematically in Fig.\ \ref{sc}, but we do not know how far it extends outside the region of convergence of the strong-coupling expansion.
 
    The existing data is at least consistent with our conjecture.  To proceed further, it will be necessary to invent and test more operators, beyond the Dirac and pseudomatter states studied so far, which might falsify (or, alternatively, support) this proposal.   We hope to report on these efforts at a later time.
    
\section{Conclusions}

 We have defined a generalization of the Wilson area law criterion, ``S$_\text{c}$ confinement,'' which is applicable to
gauge theories with matter fields in the fundamental representation,  and shown that in gauge-Higgs theories there must exist a transition between two physically distinct (S$_\text{c}$ and C) types
of confinement.  We have, in addition, suggested an alternative distinction between the Higgs and confinement phases based on custodial symmetry in the Higgs sector, and shown that this symmetry breaks spontaneously, in the special sense described above, as detected by a gauge-invariant order parameter.  Our conjecture is that the S$_\text{c}$-to-C confinement transition and the gauge-invariant symmetry-breaking transition coincide.
  
  For a more extensive presentation of the work summarized here, please see refs.\  \cite{Greensite:2017ajx,Greensite:2018mhh}.

\bibliographystyle{JHEP}
\bibliography{sym3}

....


\end{document}